\documentclass[useAMS,usenatbib]{mn2e}

\usepackage{natbib, aas_macros}
\usepackage{ulem}
\usepackage[dvips]{graphicx}
\usepackage{wrapfig}
\graphicspath{{./figs/}}
\usepackage{color}
\usepackage{amsmath}
\usepackage{amssymb}

\newcommand{\Msun}{M_{\odot}}

\newcommand{\fesc}{f_{\rm esc}}

\newcommand{\Nion}{\dot{N}_{\rm Ion}^{\gamma}}

\newcommand{\lya}{\rm {Ly{\alpha}}}

\newcommand{\Fdrag}{F_{\rm drag}}
\newcommand{\thalf}{t_{\rm half}}
\newcommand{\ttrap}{t_{\rm trap}}
\newcommand{\tlife}{t_{\rm life}}
\newcommand{\tcross}{t_{\rm cross}}
\newcommand{\kms}{\rm km\;s^{-1}}
\newcommand{\Vinit}{V_{\rm Init}}
\newcommand{\rvir}{R_{\rm vir}}

\title[Angular momentum loss of primordial gas in Ly$\alpha$ radiation field]
{Angular momentum loss of primordial gas in Ly$\alpha$ radiation field}
\author[Yajima et al.]
{Hidenobu Yajima$^{1, 2, 3}$\thanks{E-mail: yajima@roe.ac.uk (HY)} and Sadegh Khochfar$^{1}$
\\
$^{1}$ SUPA\thanks{Scottish Universities Physics Alliance}, 
Institute for Astronomy, University of Edinburgh, Royal Observatory, Edinburgh, EH9 3HJ, UK\\
$^{2}$Department of Astronomy and Astrophysics, Pennsylvania State University,
525 Davey Lab, University Park, PA 16802, USA\\
$^{3}$Institute for Gravitation and the Cosmos, The Pennsylvania State University, University Park, PA 16802, USA\\
}
 
\begin{document}

\date{Accepted ?; Received ??; in original form ???}

\pagerange{\pageref{firstpage}--\pageref{lastpage}} \pubyear{2008}

\maketitle

\label{firstpage}

%
%
\begin{abstract}

We present results on the radiation drag exerted by an isotropic and homogenous background of $\lya$ photons on neutral gas clouds orbiting within H{\sc ii}  regions around Population III stars of different masses.
The Doppler shift causes a frequency difference between photons moving in the direction of the cloud and opposite to it resulting in a net momentum loss of the cloud in the direction of motion.
We find that half of the angular momentum of gas with $v_{\theta} \lesssim 20~\kms$ near ($r \lesssim 3~\rm kpc$)  a Population III star of $120~\Msun$ at $z=20$ is lost within $\sim 10^{6}~\rm yr$. 
The radiation drag is a strong function of cloud velocity that peaks at $v \sim 20\  \kms$ reflecting the frequency dependence of the photon cross section.  
Clouds moving with velocities larger than $\sim 100~\kms$ loose their angular momentum on time scales of  $\sim 10^{8}\; \rm yr$.
At lower redshifts radiation drag becomes inefficient as the  $\lya$ photon density in H{\sc ii} regions decreases by a factor $(1+z)^{3}$ and angular momentum is lost on time scales $\gtrsim 10^{8}~\rm yr$ even for low velocity clouds.
Our results suggest that a sweet spot exists for the loss of angular momentum by radiation drag for gas clouds at $z > 10$  and with $ v \sim 20 \ \kms$. 
Comparison to dynamical friction forces acting on typical gas clouds suggest that radiation drag is the dominant effect impacting the orbit.
We propose that this effect  can suppress the formation of extended gas discs in the first galaxies and help gas accretion near galactic centres and central black holes. 

\end{abstract}

%
%

%
%
\section{Introduction}
Unraveling the evolution from Population III (Pop III) stars to the first galaxies is an open issue.    
The key to gain insight into this transition is to understand how gas is accreted onto proto-galaxies and its ability to conserve angular momentum or not. 
The latter has direct impact on the formation of possible discs \citep{Pawlik11, Romano-Diaz11}, star formation \citep{Khochfar11}  or feeding of black hole seeds \citep{DiMatteo12, Agarwal12, Agarwal13} . 

In this study, we investigate the angular momentum loss of accreted gas clouds around Pop III stars due to radiation drag by $\lya$ photons.
In an isotropic radiation field, moving gas is exerted to stronger radiation pressure along opposite sites with respect to the moving direction due to the Doppler shift. 
A similar effect on dust is classically known as the Poynting-Robertson effect \citep{Poynting1903, Robertson37}
and used to  explain angular momentum loss of dust in the solar system. 

The effect of the radiation drag on primordial gas clouds in the early universe was studied analytically \citep{Loeb93} and numerically \citep{Umemura93}.
\citet{Loeb93} and \citet{Umemura93} focused on the radiation drag exerted by the cosmic microwave background and 
suggested angular momentum of gas could efficiently be lost at $z \sim 200$, resulting in supermassive stars under the assumption that stellar source cause cosmic reionization, and that the CMB photons can efficiently be scattered by free electrons. 

However, current theoretical work shows Pop III stars form at $z \sim 15-20$ followed by the first galaxies at $z\sim 10-15$ \citep[e.g.,][]{Bromm11}.
The CMB is therefore no longer a possible source for efficient radiation drag on primordial gas.

In this paper we revisit the effects of radiation drag, in contrast to earlier work focusing on  $\lya$ radiation fields produced by Pop III stars. 
Unlike in the case for CMB photons, the number density of $\lya$ photons from Pop III stars can be much higher because they can be locally trapped within H{\sc ii} regions by scattering on their borders. 
In addition, the cross section near the line centre is quite high $\sim 10^{-14}~\rm cm^{2}$ \citep{Verhamme06}, which is $\sim 10$ orders of magnitude higher than the Thomson scattering cross section. 
As a result, the orbiting gas clouds in the $\lya$ radiation fields experience numerous scattering events  and are subject to a drag force.
However, unlike Compton scattering of continuum radiation, scattering of  $\lya$ photons is very sensitive to the velocity of the cloud. The Doppler shift of the line centre and the strong frequency dependence of the cross section may lead to an only small drag force.
In this letter, we study the effect of radiation drag on accreted gas clouds under different physical conditions such as e.g., different stellar masses of Pop III stars, redshifts, and initial velocities of infalling clouds. 

Our paper is organized as follows.
We describe our model and method to calculate the radiation drag in Section~\ref{sec:model}. 
In Section~\ref{sec:result}, we show the velocity evolution of test gas particles, and 
show the dependence on stellar mass and redshift. 
Finally, in Section~\ref{sec:summary}, we summarize our main conclusions and discuss limitations of our model.

%
%
\section{Model}
\label{sec:model}

Based on the current standard cold dark matter paradigm, halos grow via merging and accretion.  
High-density gas entering the H{\sc ii} regions around Pop III stars, which exceeds the virial radius of the halo \citep{Abel07}, can stay neutral due to self-shielding \citep[e.g.,][]{Yajima12a},  
but experience numerous scatter events with  $\lya$ photons \citep{Dijkstra12, Laursen13}. 
The critical density required for such clouds to stay neutral via self-shielding in the vicinity of the Pop III star depends on the flux of stellar radiation seen by these clouds.
Following \citet{Umemura12} we can estimate the threshold density via,
\begin{equation}
\Nion \frac{\pi r_{\rm c}^{2}}{ 4 \pi D^{2}} < \frac{4 \pi r_{\rm c}^{3}}{3} \alpha_{\rm B} n_{\rm c}^{2},
\label{eq:ion}
\end{equation}
where $\Nion$ is the ionizing photon emissivity of the star, 
and $\alpha_{\rm B}$ is the recombination coefficient to all excited levels of hydrogen \citep{Hui97}, 
$D$ is the distance between the star and the gas cloud, $n_{\rm c}$ is the hydrogen number density of the cloud and $r_{\rm c}$ its radius. 
In this work, we assume that the gas temperature is $10^{4}~\rm K$ in the H{\sc ii} region with $\alpha_{\rm B} = 2.6 \times 10^{-13}~\rm cm^{-3} \; s^{-2}$ \citep{Hui97}.
The left-side of the above equation is the maximum incident photon number, while the right-hand side is the total recombination rate in an ionized cloud. 
Using typical parameters and rearranging the threshold density for self-shielding is given by,
\begin{equation}
n_{\rm th} = 8.9 \times10^{-2}~{\rm cm^{-3}} \left( \frac{\Nion}{10^{50}~ \rm s^{-1}} \right)^{\frac{1}{2}} \left( \frac{D}{\rm 1~kpc} \right)^{-1} \left( \frac{r_{\rm c}}{\rm 100~pc} \right)^{-\frac{1}{2}}.
\label{eq:ionthresh}
\end{equation}

The above relation approximates the limiting case of an ionizing front (I-front) not being able to sweep through a neutral gas cloud. Once equilibrium between ionization and recombinations in the cloud is reached, the I-front stalls and ionization balance gives $\Nion \pi r_{\rm c}^{2} / (4 \pi D^{2}) = V_{\rm HII} \alpha_{\rm B} n_{\rm c}^{2}$, where $V_{\rm HII}$ is the volume of the ionized region. 
Taking the  volume-weighted neutral fraction $(\chi_{\rm HI})$ of the cloud as $1-(V_{\rm HII}/V_{\rm c})$ with $V_{\rm c} = 4 \pi r_{\rm c}^{3} / 3$ as the total volume of the cloud, 
the neutral fraction becomes $\chi_{\rm HI} = 1- \frac{3 \Nion}{16 \pi  \alpha_{\rm B} D^{2} r_{\rm c} n_{\rm c}^{2}}$.
A cloud of  $r_{\rm c} = 100~\rm pc$ with densities larger than $0.13~\rm cm^{-3}$ will stay mostly neutral $\chi_{\rm HI} \ge 0.5$ at a distance $D=1~\rm kpc$ from a star with $\Nion=10^{50}~\rm s^{-1}$.
This estimate is lower than in the case of ionization equilibrium under the optically thin approximation and same neutral fraction everywhere in the cloud, i.e.,
$\Nion \sigma_{\rm Ion} \chi_{\rm HI} / (4 \pi D^{2}) = \alpha_{\rm B} n_{\rm c} (1-\chi_{\rm HI})^{2}$ where $\sigma_{\rm Ion} = 6.3 \times 10^{-18}~\rm cm^{2}$ is the ionization cross section \citep{Osterbrock06}. 
Under this approxmation the above cloud becomes $\chi_{\rm HI} \ge 0.5$ at densities $ \ge 40.7~\rm cm^{-3}$.
However, in realistic situations the  ionizing flux diminishes in the cloud and the neutral fraction can be much higher than the estimation under the optically thin approximation and typically deviates from this approximation at $n_{\rm c} \gtrsim 0.01~\rm cm^{-3}$ \citep{Yajima12a, Rahmati13}, suggesting this estimate to be an upper limit on the density.  
In addition, the equation~\ref{eq:ionthresh} can be converted to a column density threshold by multiplying $2r_{\rm c}$, i.e.,
$N_{\rm th} = 5.5 \times 10^{19} ~{\rm cm^{-2}}~(\Nion/10^{50}~{\rm s^{-1}})(D/1~{\rm kpc})^{-1}(r_{\rm c}/100~{\rm pc})^{1/2}$. 
A cloud of $N_{\rm th} = 5.5 \times 10^{19} ~{\rm cm^{-2}}$  corresponds to a typical  Lyman limit system \citep{Prochaska10}, and is
mostly optically thick to external UV radiation as shown in recent cosmological simulations \citep{Faucher11, Rahmati13a}.
Hence, clouds with $n > n_{\rm th}$ are self-shielded against stellar radiation which is also supported by recent radiative-hydrodynamics simulations \citep{Susa06b, Umemura12}.
To estimate whether infalling gas clouds at these high redshifts can indeed stay neutral we consider the case of a cloud in a mini halo being accreted at $z=20$. Its mean density is  $\sim 18\pi^{2}$ times that of the background \citep{Bryan98}, or $n_{\rm H}=1.9\times10^{-7} \times (1+20)^{3} \times 18\pi^{2} = 0.3~\rm cm^{-3}$. 
This is larger than the estimated threshold assuming a Pop III star of $120~ \Msun$ as the central source, and predicts neutral infall.
Further support comes from simulations by \citet{Abel07} who show that a gas cloud accreted along a cosmic filament does survive photoevaporation even at a distance of 50 pc from the star \citep[see also][]{Umemura12}. 
On the other hand, the threshold density is sensitive to the separation to the star. 
If clouds enter the  virial radius,  within the life time of stars ($\sim 3$ Myr), for a virial radius of 
$D \sim 70$ pc (corresponding to a dark matter halo of $\sim 10^{5}~\Msun$ at $z=20$) the threshold density gets $n_{\rm th} = 1.3~\rm cm^{-3}$ which is larger than the mean density of the halo.
It is therefore likely that some fraction of the infalling clouds, in particular those that come closest to the star on a short time scale, are destroyed due to the photoevaporation \citep[see also][]{Hasegawa09a, Susa09}. 

Even after the death of the central star in a H{\sc ii} region, $\lya$ photons are produced over a recombination timescale ($\tau \sim 6 \times10^{7}~\rm yr$ at z=20), and  can be trapped in the residual H{\sc ii} region. 
As well as $\lya$ photons,
ionizing photons are also emitted by recombination from a halo. 
The total power of ionizing photons from a halo by recombination is similar to that by stellar sources \citep[e.g.,][]{Raicevic13} leaving infalling clouds neutral due to self-shielding.
The recombination time scale of gas in haloes is much smaller than that of the inter-galactic medium (IGM), $\tau \sim 6\times10^{5}$ yr terminating ionizing photons from a halo at almost the same time when the central star dies. Although the production of ionizing photons from the IGM is continuing its emissivity is small due to its low density.  At this stage, even lower-density accreted gas can stay neutral because there is no strong ionizing photon contribution and interact with the abundant $\lya$ photons.

\begin{figure}
\begin{center}
\includegraphics[scale=0.3]{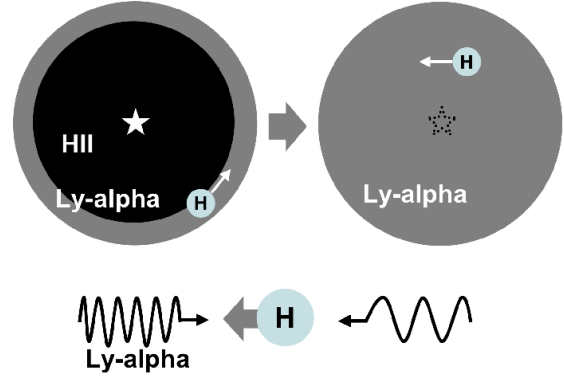}
\caption{
Schematic view of our model. 
The high-density region of $\lya$ photons is generated by recombinations in the H{\sc ii} region. 
Gas can feel higher $\lya$ radiation pressure to the opposite of the moving direction due to the Doppler shift, resulting in the radiation drag. 
}
\label{fig:img}
\end{center}
\end{figure}

 $\lya$ photons interacting along the direction of motion have larger energies than those  opposite to it due to the Doppler shift. 
 This  leads to a net momentum loss and drag force in the direction of motion of $\Delta P \sim 2\times h\nu_{0}(v/c^2)$ , where $\nu_{0}=2.466\times10^{15}~\rm Hz$ is the frequency of the line centre. 
 Figure~\ref{fig:img} shows the schematic view of our model. 
The drag force is proportional to the number density of $\lya$ photons locked up within the H{\sc ii} region. In the following paragraphs we derive estimates for the number density of $\lya$ photons.

We estimate the size of H{\sc ii} regions assuming a uniform inter-galactic medium (IGM) around Pop III stars and neglect the clumpiness of the IGM which has been shown in numerical simulations to be small and $\lesssim 3$ \citep{Pawlik09b, Paardekooper13}.
This leads to an estimate of the size of the H{\sc ii} region in equilibrium based on the Str{\"{o}}mgren sphere analysis \citep{Stromgren39} : 
\begin{equation}
R_{\rm S} = \left( \frac{3\dot{N}_{\rm Ion}^{\gamma}\fesc}{4\pi \alpha_{\rm B} n_{\rm H}^{2}(z)} \right)^{\frac{1}{3}},
\end{equation}
where $R_{\rm S}$ is the radius of the Str{\"{o}}mgren sphere, $\Nion$ is the ionizing photon emissivity of stars, $\fesc$ is the escape fraction of ionizing photons from the halo.
For small haloes hosting Pop III stars, most of the gas can easily evaporate due to photo-ionization and a large fraction of ionizing photon can escape \citep{Whalen04, Kitayama04}.
 Note however, that estimates for $\fesc$ can vary significantly 
with halo and stellar mass \citep{Kitayama04, Yoshida07}, and are somewhat dependent on the resolution of the simulation \citep{Yajima11, Rahmati13}.
Here we assume $\fesc = 0.5$ which is supported by recent simulations \citep{Abel07, Paardekooper13}.
The recombination time scale is of order $10^{7}-10^{8}~\rm yr$ and as such longer than the life time of massive Pop III stars resulting in the ionization front being within $R_{\rm S}$ during the life time of the star \citep{Spitzer78}, 
\begin{equation}
R_{\rm I} = R_{\rm S} \left( 1 - {\rm exp}(-t_{\rm life}/\tau_{\rm rec})\right)^{1/3},
\end{equation}
where $t_{\rm life}$ is life time of the star, $\tau_{\rm rec} \sim 1/\alpha_{\rm B} n_{\rm H}$ is the recombination time-scale, and $ R_{\rm I}$ is the radius of the ionization front. 
About 0.68 of ionizing photons emitted from stars are converted to $\lya$ photons via recombinations \citep{Osterbrock06}. 
A large fraction of these $\lya$ photons is scattered at the outer H{\sc i} layer \citep{Verhamme06, Dijkstra08}  and locked inside the H{\sc ii} region, leading to a significant increase in their number density \citep{Yajima12g}. 
We estimate the total number of $\lya$ photons in H{\sc ii} regions as
\begin{equation}\label{nlya}
N_{\lya} \sim 0.68 \times (1 - \fesc) \times \Nion \times \ttrap,
\end{equation}
where $t_{\rm trap}$ is the trapping time of $\lya$ photons in the H{\sc ii} region. 
Here, as a fiducial model, we assume for simplicity that $\ttrap = \tlife$.
This model implies that absorbed ionizing photons in haloes convert to $\lya$ photons via the recombination process and are then stocked in the ionized region. 
The drag force $\Fdrag$ is proportional to the number density of $\lya$ photons 
and thus  $\Fdrag \propto (1-\fesc)/\fesc$ because the ionized regions become smaller as $V_{\rm HII} \propto \fesc$ 
and the total $\lya$ photon number increases with $N_{\lya} \propto (1-\fesc)$.
Hence,  $\Fdrag$ increases as the $\fesc$ decreases, while the volume of the trapped $\lya$ radiation field becomes smaller.
The frequency of $\lya$ photons changes during each scattering event. 
Typically after an average of $N_{\rm scat} \sim 0.9 \tau_{0}$ scattering events \citep{Harrington73}, the $\lya$ photon frequency changes to frequencies with low cross sections and they escape from the previously optically thick medium. 
In the case of outflowing gas like the IGM,  only photons at longer wavelength than $\nu_{0}$ can travel for long distances \citep{Loeb99, Laursen09a, Yajima12b}.
On the other hand, for circumgalactic gas  that is infalling only photons at shorter wavelengths can escape from the local region \citep{Yajima12f}.
If we consider an outflow paralleling the Hubble flow, the optical depth for the IGM is $\tau_{0} \sim10^{6}$. 
Under such circumstances $\lya$ photons are efficiently scattered at the outer H{\sc i} layer \citep{Verhamme06, Dijkstra08, Yajima13b}. 
The trapping time of $\lya$ photons ($\ttrap$) in H{\sc ii} regions can be much longer than the life time of Pop III stars $\sim 10^{6}~\rm yr$ \citep{Schaerer02} 
by $\ttrap \sim 0.9 \tau_{0} \times 2R_{\rm I} / {\rm c} \gtrsim 10^{10}~\rm yr$, here we use $R_{\rm I} \sim 2.9 ~\rm kpc$ representative for a Pop III star with 120 $\Msun$ at $z=20$.
Hence, all emitted $\lya$ photons from haloes can be locked within the H{\sc ii} region. 
This estimate is an upper limit, as some fraction can enter into the H{\sc i} layer and escape.
 Please note that even after a star dies, the recombination processes in the residual H{\sc ii} region keeps producing $\lya$ and ionizing photons.
The total number of $\lya$ photons continues increase during this time due to the trapping effect, while ionizing photons are destroyed via absorption by hydrogen and recombination to excited levels. 
In this work we use a constant value for the density of $\lya$ photons based on Eq.\ref{nlya}.
However, the trapping time becomes shorter in the recombining IGM. 
\citet{Adams75} derived the trapping time in spherical uniform  and plane-parallel clouds.  
He showed $t_{\rm trap} \sim 15 t_{\rm cross}$ for $3 <{\rm log}\tau_{0} < 6$, where $\tau_{0}$ is the optical depth of the $\lya$ line centre and $t_{\rm cross}$ is the crossing time of photons through a system of size $L$, i.e., $t_{\rm cross} = L / \rm c$. 
To account for this we also study cases with lower photon densities with $\ttrap = 15\tcross$ with $\tcross = 2R_{\rm I} / \rm c$.
 The number density of $\lya$ photons in the H{\sc ii} is then estimated by 
$n_{\lya} = N_{\lya} / V_{\rm HII} = 3 N_{\lya} / 4\pi R_{\rm I}^{3}$. 

In our model we assumes a spherical symmetric H{\sc ii} region. Depending on the environment and physical situation, geometries in general can vary from highly non-symmetric \citep{Abel07} to spherical \citep{Yoshida07}.  
However, the number density of $\lya$ photons is dominated by the ionizing photon emissivity from the central star and the volume of H{\sc ii} region as explained above. 
It is not sensitive to the detailed shape of the H{\sc ii} region. 
We further assume a homogeneous $\lya$ radiation field in the H{\sc ii} region.
In practice, since $\lya$ photons are emitted from high-density gas within the halo, 
the $\lya$ photon density increases towards the halo.
However, due to the photon trapping effect and the associated scattering of Lyman-alpha photons into random directions within the H{\sc ii} region,   
the cumulative number of $\lya$ photons is much higher than the in situ emission from the halo and the $\lya$ radiation field becomes highly homogeneous. 

For conveniency, in what follows we introduce a variable $x \equiv (\nu - \nu_{0})/\Delta\nu_{\rm D}$, where  $\Delta\nu_{\rm D} = [2k_{\rm B}T/(m_{\rm p}c^{2})]^{1/2}\nu_{0}$. 
The intensity can then be expressed as,
\begin{equation}
I(x) = I_{0} H(a, x), 
\end{equation}
where
\begin{equation}
I_{0} = \frac{n_{\lya} {\rm c}}{ 4\pi \int dx H(a,x)},
\end{equation}
 and $a=\Delta\nu_{\rm L}/(2\Delta\nu_{\rm D})$ is the relative line width with the natural line width $\Delta\nu_{\rm L}=9.936\times10^{7}~\rm Hz$,
and $H(a,x)$ is the Voigt function \citep{Verhamme06},
\begin{equation}
H(a, x) = \frac{a}{\pi} \int_{-\infty}^{+\infty} \frac{e^{-y^{2}}}{(x-y)^{2} + a^{2}} dy
\sim \begin{cases}
e^{-x^{2}} & {\rm if} ~{\rm |x| < x_{\rm c}}\\
\frac{a}{\sqrt{\pi} x^{2}}
& {\rm if} ~{\rm |x| > x_{\rm c}}, 
\end{cases}
\end{equation}
where $x_{\rm c}$ is the boundary frequency between a central resonant core and 
 the power-law ``damping wings''. 
 We use the fitting formula by \citet{Tasitsiomi06} which can reproduce the Voigt function smoothly even at $x \sim x_{\rm c}$.
 During each scattering event, the frequency of $\lya$ photons changes. However, when the optical depth of the IGM is high, a very small fraction of $\lya$ photons at the wing part of the distribution can escape from local regions. Hence, the line profile in H{\sc ii} region stays a Voigt function. 
 Furthermore, the propagation of I-front does not accompany high-velocity motion of gas \citep{Yoshida07}. 
 Therefore, the line profile of $\lya$ photons is not significantly affected by the I-front propagation. 
 
In the rest frame of the gas, the radiation field is not isotropic due to the Doppler shift of photons,  resulting in a net drag force. 
By considering the anisotropic radiation pressure \citep{Dijkstra08}, the radiation drag in the $\lya$ radiation field is estimated 
as
\begin{equation}
\Fdrag = \frac{4\pi}{\rm c} \int dx \sigma(x) K(x),
\end{equation}
where
\begin{equation}
K(x) = \int d\mu\; \mu I(x),
\end{equation}
with $\mu \equiv {\rm cos}\theta$.
The cross section is estimated by 
$\sigma(x) = 1.04\times10^{-13} T_{4}^{-1/2} H(a,x)/\sqrt{\pi}$ \citep{Verhamme06}.
%

The change in velocity for a gas cloud due to this drag is then
$dV / dt = - \Fdrag$.
In what follows we will show results for the velocity evolution of gas clouds around a Pop III star of $120~\Msun$ at $z=20$ as a fiducial case, and study cases with different masses and redshifts. For simplicity we will refer to velocity and angular momentum below and not separately distinguish between tangential and radial velocities, and momentum and angular momentum, respectively, as the $\lya$  field is isotropic and homogenous.

%
%

\section{Results}
\label{sec:result}

\begin{figure}
\begin{center}
\includegraphics[scale=0.4]{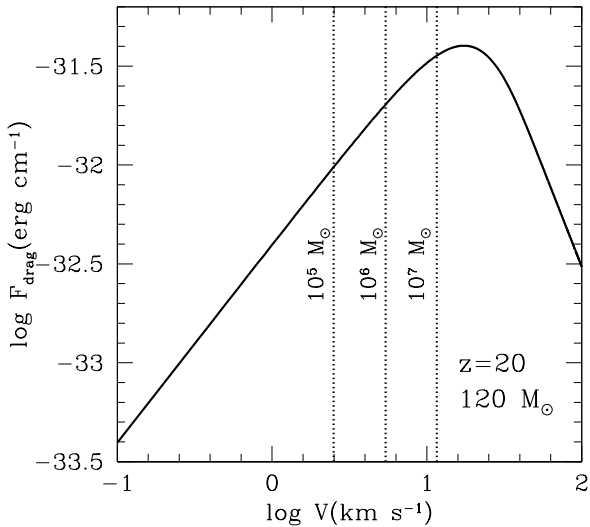}
\caption{
Radiation drag as a function of velocity of gas clouds around a Pop III star of $120~\Msun$ at $z=20$. Dotted lines show virial velocities  of $10^{5}, 10^{6}$ and $10^{7}~\Msun$ halos at $z=20$. }
\label{fig:frad_v}
\end{center}
\end{figure}

Figure~\ref{fig:frad_v} shows $\Fdrag$ as a function of gas velocity. 
The drag force $\Fdrag$ is roughly proportional to the cross section and energy difference due to the  Doppler shift, both of which are functions of the relative velocity
with respect to the background radiation field.
The energy difference of photons by Doppler shift is proportional to $\sim x$, $\Delta E \propto x$. 
Hence, if $x \ll 1$,  and $\sigma \propto e^{-x^{2}} \sim 1 - x^{2}$, as a result, $\Fdrag \propto x \propto v$. 
On the other hand, if $x \gg 1$, $\sigma \propto x^{-2}$, hence $\Fdrag \propto x^{-1} \propto v^{-1}$.
Therefore, the $\lya$ radiation drag has a peak near $x \sim 1$. 
In comparison, radiation drag due to the Compton scattering of continuum radiation is simply proportional to the gas velocity \citep{Umemura93}.
Furthermore, although the absolute value of $\Fdrag$ changes with stellar mass and redshift due to the change in the  number density of $\lya$ photons, the functional shape does not. 
The shape depends only  on the temperature of the gas from which $\lya$ photons are emitted. 
In this work, we assume $T=10^{4}~\rm K$, as a result the peak is at $V = 17~\kms$. 
With increasing temperature, the peak position moves to slightly larger velocities. 
However, the dependence on temperature is not significant in the range of $10^{4} \le T \le 3\times10^{4}~\rm K$ which is the typical temperature of H{\sc ii} region. 
For $T=10^{4}~\rm K$ and $n_{\rm Ly\alpha}=2.5\times10^{-3}~\rm cm^{-3}$ (the case of $M_{\rm PopIII}=120~\Msun$ at $z=20$), 
we can roughly fit the drag force by $\Fdrag = 0.4\times10^{-32} {\rm erg\;cm^{-1}}(v/1~\kms)$ if $v < 17~\kms$,
and $\Fdrag = 0.4\times10^{-31} {\rm erg\;cm^{-1}}(v/20~\kms)^{-1}$ if $v > 17~\kms$. For different number densities $n_{\rm Ly\alpha}$ the drag force can be estimated scaling these fits linearly with the number density.

\begin{figure}
\begin{center}
\includegraphics[scale=0.45]{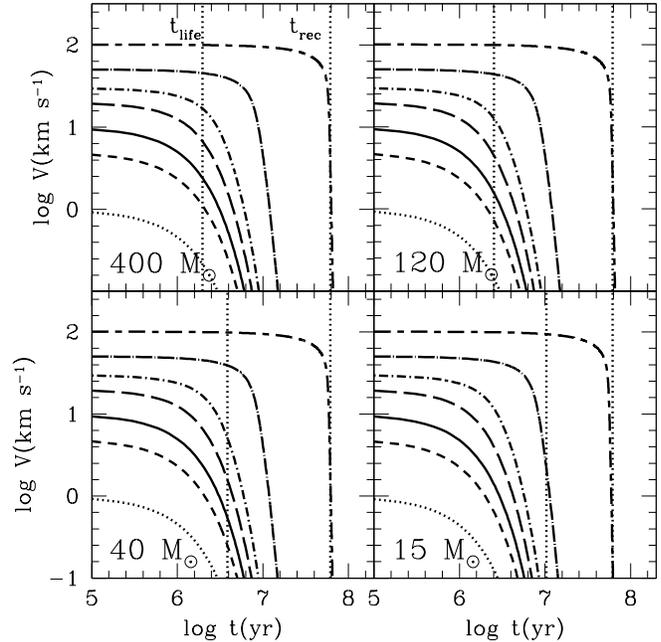}
\caption{
Time evolution of the gas velocity of a neutral gas cloud in a $\lya$ radiation field generated around Pop III stars at $z=20$.
Different line types represent different initial velocities of the cloud.
Vertical dotted lines are life times of stars and a recombination time scale at $z=20$.}
\label{fig:v_mass}
\end{center}
\end{figure}

The initial mass function of Pop III stars is still under debate, it likely depends on the environment of the formation site.
Recently, \citet{Hirano13} carried out a large set of cosmological hydrodynamics simulations following 100 Pop III stars forming, and suggested that the mass of Pop III stars might range from $ \sim 10 $ to $\sim 1000~\Msun$.
To account for this uncertainty, we here calculate  $\Fdrag$ around Pop III stars for  a range of  masses with the ionizing photon emissivities and life times of Pop III stars taken from \citet{Schaerer02}.
The velocity evolution of gas clouds around Pop III stars of $15, 40, 120$ and $400~\Msun$ at $z=20$ are shown in Figure~\ref{fig:v_mass}.
Different line types represent different initial velocities of clouds. 
As shown in the Figure, there is no clear difference in the velocity evolution for different stellar masses. 
Since luminosity is linearly proportional to stellar mass and the effective temperature for the stars considered here is almost constant at  $T\sim10^{5}~\rm K$ \citep{Bromm01}, the ionizing photon emissivity linearly increases with stellar mass $\Nion \propto M_{\rm PopIII}$.
As shown in equation (1) and (2), $V_{\rm HII} = R_{\rm S}^{3} [1-{\rm exp}(-\tlife/\tau_{\rm rec})] \sim R_{\rm S}^{3} (\tlife/\tau_{\rm rec}) \propto \Nion \tlife$.   
Therefore,  $n_{\lya} \propto \Nion \tlife / V_{\rm HII} = \rm const$ and hence the $\lya$ radiation drag does not depend on stellar mass. 

Clouds with $\Vinit=10~\kms$ loose half of their angular momentum at $t = 1.0\times10^{6}~\rm yr$, and move with $V = 0.1~\kms$ at $t=6.0\times10^{6}~\rm yr$. 
Gas clouds with $\Vinit=20$ and $30~\kms$  are beyond the peak in  $\Fdrag$ but have somewhat higher drag forces acting on them than those at $V=10~\kms$.
These clouds pass through the peak of $\Fdrag$ in their velocity evolution, and hence the time-averaged $\Fdrag$ is higher than that of clouds starting out at $\Vinit=10~\kms$.
As a consequence, although their initial angular momenta are higher, the time scale required to reach $V=0.1~\kms$ is only slightly larger to that of $\Vinit=10~\kms$.
For gas with $\Vinit=100~\kms$, the velocity evolution is very slow, and becomes $V=0.1~\kms$ at $t = 6.6\times10^{7}~\rm yr$  which is comparable to the recombination time scale at $z=20$.  
After recombination of the IGM, the mean free path of $\lya$ photons becomes much shorter, and their frequency can quickly change to the red wing part, resulting in an escape from the local region and a drop in the number density of $\lya$ photons. 
In effect only gas with  $\Vinit \lesssim 100~\kms$ loose a significant fraction of its angular momentum.

The circular velocities of dark matter haloes of $10^{6}$ and $10^{7}~\Msun$ at $z=20$ are corresponding to $5.4$ and $11.6~\kms$ respectively. 
The peak velocity of $\Fdrag$ is corresponding to haloes with circular velocity of $3 \times 10^{7}~\Msun$ at $z=20$ and $9\times10^{7}~\Msun$ at $z=10$.
In the current standard scenario, Pop III stars form in mini-halos of $\sim 10^{6}~\Msun$ at $z\sim20$ \citep[e.g.,][]{Yoshida08}, subsequently evolving into the first galaxies with $\sim 10^{8}~\Msun$ at $z \sim 10$ \citep[e.g.,][]{Wise12a} via gas accretion and mergers. 
Given that the time scale associated with the loss of angular momentum is shorter than the dynamical time scale at $z=20$ we expect this effect to impact the growth from the first stars to the first galaxies.

Please note that we here assume a uniform IGM density field in which the photon drag takes place. This is justified outside the virial radius of the halo. We have neglected the interaction within the virial radius which is subject to  complex geometry and physical state of the gas. However, the size of the H{\sc ii} region is $\sim 10$ times larger than the typical virial radius ($V_{\rm Halo}/V_{\rm HII} \sim 10^{-3}$), and most of the angular momentum will be lost during the phase outside the virial radius.

\begin{figure}
\begin{center}
\includegraphics[scale=0.4]{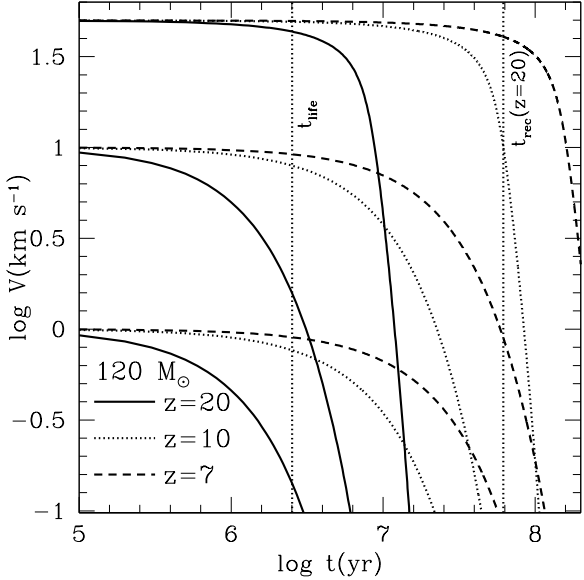}
\caption{
Time evolution of  gas cloud velocities in $\lya$ radiation field surrounding  a Pop III star of $120~\Msun$ at different redshifts. 
Vertical dotted lines are a life time of star and a recombination time scale at $z=20$.
}
\label{fig:v_z}
\end{center}
\end{figure}

As well as having a wide range of stellar mass, Pop III stars can form at various redshifts \citep{Wise12a}.
Figure~\ref{fig:v_z} shows the velocity evolution at different redshifts. 
With decreasing redshift, $\Fdrag$ decreases, resulting in slower velocity change. 
The Hydrogen number density of the IGM increases with redshift as $(1+z)^{3}$, resulting in $R_{\rm S} \propto (1+z)^{-2}$. 
In addition, with decreasing gas density, the recombination time scale becomes larger, $\tau_{\rm rec} \sim 1/(\alpha_{\rm B}n_{\rm H}(z))$ leading to 
$R_{\rm I} = R_{\rm S} (1-{\rm exp}(-\tlife/\tau_{\rm rec}))^{1/3} \sim R_{\rm S}(\tlife/\tau_{\rm rec})^{1/3} \propto n_{\rm H}^{-1/3} \propto (1+z)^{-1}$.
As a result, $\Fdrag$ changes with redshift as  $(1+z)^{3}$ because of $n_{\lya} \propto V_{\rm HII}^{-1}$. 
At $z=7$, it takes $1.2 \times 10^{8}~\rm yr$ until a cloud with $\Vinit = 10~\kms$ slows down to $0.1~\kms$, which is longer than the recombination time scale. 
This effectively limits the  $\lya$ photon drag to high redshifts $(z > 7)$ even for $\Vinit \lesssim 10~\kms$.

In this work, our model assumes H{\sc ii} regions are surrounded by a neutral IGM, however, H{\sc ii} bubbles in the IGM can overlap with each other at later redshifts.
In this  case the distance that $\lya$ photons travel to residual H{\sc i} boundaries is much larger and the scattering cross section is lower due to the Hubble flow.
For example, \citet{Iliev06a} show the size of overlapping H{\sc ii} bubbles around galaxies can be $\gtrsim 10~\rm Mpc$. 
Hence, as H{\sc ii} bubbles overlap, the relative velocity of the H{\sc i} boundary is much larger than the typical velocity width of the $\lya$ line. At this point most of $\lya$ photons escape from the large H{\sc ii} bubble without scattering, and the $\lya$ radiation drag is no longer effective.

\begin{figure}
\begin{center}
\includegraphics[scale=0.45]{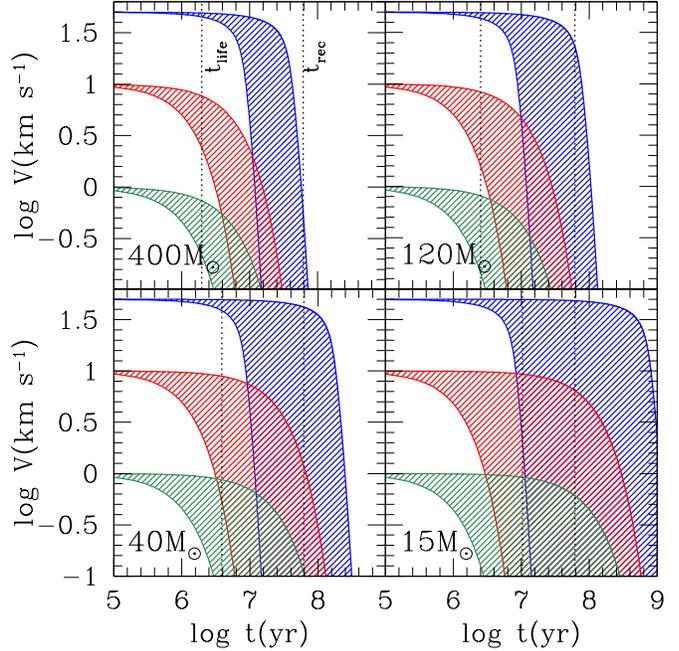}
\caption{
Time evolution of gas cloud  velocity in a  $\lya$ radiation field around Pop III stars at $z=20$.
The upper and lower values in the shaded regions correspond to the cases of $t_{\rm trap} = 15\times t_{\rm cross}$ and 
$t_{\rm trap} = t_{\rm life}$ respectively, where $t_{\rm trap}$ is the trapping time of $\lya$ photons, $t_{\rm cross}$ is the crossing time 
of the system, and $t_{\rm life}$ is life time of the host star. 
Dotted lines are life times of stars and a recombination time scale at $z=20$.
}
\label{fig:v_shade}
\end{center}
\end{figure}

To gain further insight into the effect of the trapping time onto the $\lya$ photon drag, we consider that 
after the death of stars, ionized hydrogen starts to recombine within H{\sc ii} regions.  Once the neutral fraction increases,  the mean free path of $\lya$ photons become shorter as a result of which they can quickly move to energies in the wing of the line profile (Eq. 7) and escape, resulting in smaller trapping times. 
\citet{Adams75} showed the mean trapping times from optically-thick clouds with uniform density are $\ttrap \sim 15\tcross$. 
We investigate the impact of the different trapping time on the momentum loss using $\ttrap = 15\tcross$ with $\tcross = 2R_{\rm I} / \rm c$. 
Figure~\ref{fig:v_shade} shows the velocity evolution of gas around Pop III stars at $z=20$. 
The boundaries of the shaded regions indicates the velocity evolutions with $t_{\rm trap} = t_{\rm life}$ and $t_{\rm trap} = 15 t_{\rm cross}$.
In the case of $120~\Msun$, $\Fdrag$ is lower than that of the fiducial model with $t_{\rm trap} =  t_{\rm life}$ by a factor $\tlife / (15\tcross) = 8.9$. 
As a result the velocity of the gas reduces only slowly.
The velocity of gas clouds with $\Vinit=10~\kms$ becomes half at $t=0.9\times10^{7}~\rm yr$ and $0.1~\kms$ at $t=5.6\times10^{7}~\rm yr$ for $120~\Msun$.
For high-velocity gas of $\Vinit=50~\kms$,  the velocity evolution is even slower, and the velocity becomes $0.1~\kms$ at $t=1.3\times10^{8}~\rm yr$.

In addition, as shown in Figure \ref{fig:v_shade}, the velocity evolution in the case of $t_{\rm trap} = 15 t_{\rm cross}$ significantly depends on stellar mass.
With decreasing stellar mass, $\Fdrag$ becomes lower and hence the velocity evolution gets slower. 
This is because, $R_{\rm I} \propto N_{\rm Ion}^{1/3}$, and 
$V_{\rm HII} \propto R_{\rm I}^{3}$ and $\ttrap \propto R_{\rm I}$, 
as a result, $\Fdrag \propto \Nion R_{\rm I} / V_{\rm HII}= M_{\rm PopIII}^{1/3}$. 
For a $400~\Msun$ star a gas cloud with $\Vinit=10~\kms$ looses half of initial angular momentum at $t = 5.0\times10^{6}~\rm yr$
and becomes $V=0.1~\kms$ at $t=3.0\times10^{7}~\rm yr$.
On the other hand, for a $15~\Msun$ star, it takes $9.5\times10^{7}~\rm yr$ until the initial velocity of $10~\kms$ reduces to half that,
and $5.8\times10^{8}~\rm yr$ until it is $0.1~\kms$.

\begin{figure}
\begin{center}
\includegraphics[scale=0.4]{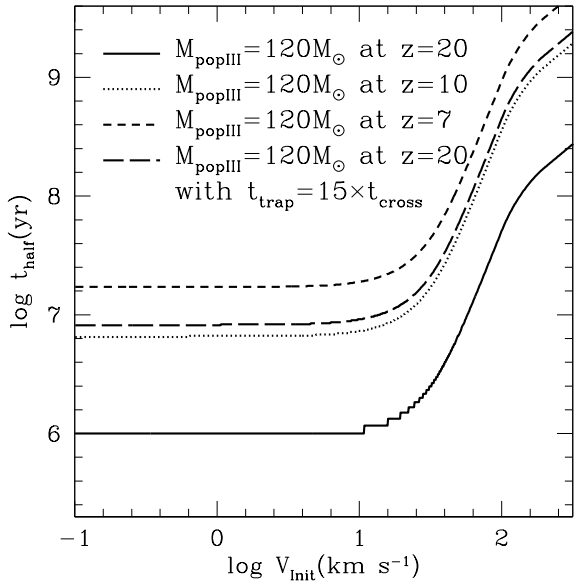}
\caption{
Time scale until the initial velocity becomes half. 
Different line types are corresponding to different situations.
Solid, dot and dash line represent the models for $M_{\rm PopIII}=120~\Msun$ at $z=20, 10$ and $7$ respectively.
Long dash line shows the model for $M_{\rm PopIII}=120~\Msun$ at $z=20$ with a photon trapping time of 15 times the crossing time. 
}
\label{fig:half}
\end{center}
\end{figure}

Figure~\ref{fig:half} shows the time by which gas has lost half of its initial velocity ($\thalf$).
Depending on the situations,  $\thalf$ changes, and ranges between $ \sim 1\times10^{6}$  to $\sim 2\times 10^{7}$ yr for $V \lesssim 20~\rm \kms$.
For $V \lesssim 20~\rm \kms$, $\thalf$ is almost constant,
because $\Fdrag$ is roughly proportional to $V$. 
For $\Vinit > 20~\kms$, $\thalf$ steeply increases with the initial velocity,
it becomes $\sim 1~\rm Gyr$ at $\Vinit \gtrsim 100~\kms$ and becomes larger than the recombination time scale. 
After recombination, $\lya$ photons can quickly escape from local regions, resulting in a smaller trapping time.  
Therefore, for $\Vinit \gtrsim 100~\kms$, our estimation of $\Fdrag$ at $t > t_{\rm rec}$ should be considered an upper limit, and the time scale should be longer.

%
%


%
%

\section{DISCUSSION \& SUMMARY}
\label{sec:summary}

In this paper, we have investigated the effect of radiation drag due to $\lya$ scattering on accreted neutral gas clouds  around  first stars. 
We find that half of the angular momentum of gas with $v_{\rm c} \lesssim 10~\rm km\;s^{-1}$ near a Pop III star of $120~\Msun$ at $z=20$ is lost within $\sim 10^{6}~\rm yr$. 
Due to the sensitivity of the scattering cross section to gas velocities, the $\lya$ radiation drag ($\Fdrag$) declines strongly for gas clouds with velocities  $\gtrsim 20~\kms$. 
For initial gas cloud velocities of $100~\kms$, more than half of the angular momentum can be kept for $5.1\times 10^{7}~\rm yr$ against the radiation drag.
For our fiducial model with $\ttrap=\tlife$, $\Fdrag$ does not depend on stellar mass. 
On the other hand, $\Fdrag$ is sensitive to redshift. At decreasing redshifts, due to larger sizes of H{\sc ii} regions, the number density of $\lya$ photon decreases.
Consequently, $\Fdrag$ decreases with $(1+z)^{3}$. 
We find that only at redshifts $z > 10$, the $\lya$ radiation drag suppress circular motion of accreted gas efficiently, while it is not likely to block infalling stream motion like  cold accretion with $V \gtrsim 100~\kms$ \citep{Yajima12f}. 
The radiation drag time-scale is much shorter than that of the dynamical friction in gas medium \citep{Ostriker99} and the dynamical time scale at $z \sim 20$.
Hence, this effect dominates the angular momentum transport and
suppress the formation of large scale gas discs around the first objects like Pop III stars and seed black holes, and enhance the gas accretion towards them.
Note that, however, if the clouds enter the virial radius before the stars die, they are most likely destroyed due to photoevaporation and no accretion onto the central object will occur. 

Here, we focused on the gas accretion onto mini haloes in which Pop III stars formed. 
Similar $\lya$ radiation drag may occur around the first galaxies at later times. 
Since the typical formation epoch is later than that of Pop III stars, the $\lya$ photon density and $\Fdrag$ are smaller.
However, recent simulations have shown that the escape fraction of ionizing photons decreases with halo mass \citep{Yajima11, Yajima12d, Paardekooper13}.
Smaller escape fractions result in compact H{\sc ii} regions and high-density $\lya$ photon fields, resulting in higher $\Fdrag$.
In addition, continuous star formation keeps the production of $\lya$ photons ongoing and allows the radiation drag to occur for a longer time.
Hence, the situation with respect to the $\lya$ radiation drag around first galaxies is not certain.

After the death of stars, the central region in the H{\sc ii} recombines fast, resulting in a  H{\sc ii} shell in between a central H{\sc i} region and an outer H{\sc i} border. 
The shell becomes smaller as the central region continues to recombine while the number of $\lya$ photons in the shell stays roughly constant. The radiation drag increases during this period in the shell, until it has recombined or the $\lya$ photons have escaped.

While the assumption of a uniform IGM density is fair, the gas within the virial radius is highly inhomogeneous \citep{Abel07}. This limits the validity of our model from the edge of the H{\sc ii} region ($\sim$ a few kpc) to the virial radius ($\sim$ a few hundreds pc). However, as we show above this is the dominant region in which angular momentum loss occurs.  The evolution inside the halo is complex as e.g.  shock waves exist with  $\gtrsim 10~\kms$  \citep{Abel07, Yoshida07}. These are generated in tandem with the I-front.
At first, due to the high-density gas around the star, the I-front is D-type and slowly propagates with the shock.
Then, reaching  low-density regions, the I-front changes to R-type and quickly expands, while the shock is left behind stagnating at a propagation velocity of $\gtrsim 10~\kms$.
As a result, the shock is at around the virial radius when the central star dies \citep{Yoshida07}.
Once the clump reaches the virial radius, 
it is likely to interact with the shock and be compressed and heated up.
As a result, the accreted gas may be destroyed or survive by efficient cooling emission \citep{Dekel06}.
This evolution of the accreted gas at $\lesssim \rvir$ is out of the scope of our simple model in this paper.

In this work, as a first step, we use an analytic model to estimate the $\lya$ radiation drag by estimating the upper limit of the possible $\lya$ radiation field. It is important to note, that in our model trapping of $\lya$ photons is key. Once e.g. supernovae efficiently blow out gas, $\lya$ photons are able to escape the local region and the radiation drag quickly becomes unimportant. 
In a follow-up study we will investigate the $\lya$ radiation drag in a more realistic set-up including a self-consistent time evolution within  hydrodynamics simulations.

%
%
\section*{Acknowledgments}
We are grateful to Y. Li and M. Umemura for valuable discussion and comments.
We thank the anonymous referee for useful comments.

%
%




\label{lastpage}

\end{document}